\documentclass{ws-procs975x65}

\def\lsim{\lower.5ex\hbox{$\; \buildrel < \over \sim \;$}}
\def\gsim{\lower.5ex\hbox{$\; \buildrel > \over \sim \;$}}

\begin{document}

\title{Photometric Evidence of Bullets in SS433 Jets}
\author{S. K. CHAKRABARTI\footnote{\uppercase{A}lso at \uppercase{C}entre for \uppercase{S}pace \uppercase{P}hysics, \uppercase{C}halantika 43, \uppercase{G}aria \uppercase{S}tation \uppercase{R}d.,
\uppercase{K}olkata 700084} ~  and ~ A. NANDI\footnote{\uppercase{W}ork partially
supported by \uppercase{D}epartment of \uppercase{S}cience and \uppercase{T}echnology}}
\address{S.N. Bose Nat'l Centre for Basic Sciences, JD-Block, Salt Lake, Kolkata 700098\\
E-mail: chakraba@bose.res.in and anuj@bose.res.in}
\author{S. PAL\footnote{\uppercase{W}ork partially
supported by \uppercase{C}ouncil of \uppercase{S}cienctific and \uppercase{I}ndustrial \uppercase{R}search}}
\address{Centre for Space Physics, Chalantika 43, Garia Station Rd., Kolkata 700084\\
E-mail:space\_phys@vsnl.com}
\author{B.G. ANANDARAO and SOUMEN MONDAL}
\address{Physical Research Laboratory, Navarangapura, Ahmedabad, 380009, India\\
E-mail: anand@prl.ernet.in and soumen@prl.ernet.in}

\maketitle

\abstracts{ We report the photometric evidence of bullet like features in 
SS433 in X-rays, Infra-red and Radio bands through a  multi-wavelength
campaign.}

\noindent To be Published in the proceedings of the 10th Marcel Grossman Meeting (World Scientific Co., Singapore),
Ed. R. Ruffini et al.

\section{Introduction}

The enigmatic micro-quasar SS433 showed evidence of bullet-like 
ejection in optical band (e.g., Vermeulen, 1993).  Chakrabarti
et al (2002) presented a mechanism to produce 
quasi-regular bullets. These bullets would be ejected from X-ray
emitting region and propagate through optical, infra-red 
($\sim 10^{13-14}$cm) and finally to radio emitting region at 
$\gsim 10^{15}$cm (roughly the distance covered in a day 
with $v\sim v_{jet}$) or so. Thus if the object is in a low 
or quiescence state, each individual bullet flaring and dying 
away in a few minutes time scale, should be observable not 
only in optical wavelengths (Margon, 1984) but also in all 
the wavelengths, including X-ray, IR and radio emitting regions. 
Here, we report the results of a multiwavelength campaign in 
X-ray, infra-red (IR) and radio observations of September, 2002.  
Our main results indicate that there are considerable variations 
in the timescale of minutes in all the wavelengths. 
Detailed results are in Chakrabarti et al. (2003).

\section{Observation and Data Reduction}

Radio observation was carried out with Giant Meter Radio Telescope (GMRT) at
1280MHz (bandwidth $16$ MHz). The data is binned at every $16$ seconds.
AIPS package was used to reduce the data. Infrared observation was made 
using Physical Research Laboratory (PRL) 1.2m Mt. Abu infrared 
telescope equipped with Near-Infrared Camera and Spectrograph (NICMOS). 
The filters used were standard J,H and K$^\prime$ bands.  X-ray observation 
was carried out using Proportional Counter Array (PCA) on board RXTE 
satellite. The data reduction and analysis was performed using software 
(LHEASOFT) FTOOLS 5.1 and XSPEC 11.1. We extract light curves from the 
XTE/PCA Science Data of GoodXenon mode. We also extracted energy spectra 
from PCA Standard2 data in the energy range 3-25keV.

\section{Results on short time-scale variabilities}

\begin{figure}[t] 
\vskip 0.0cm
\centerline{\epsfxsize=4.1in\epsfbox{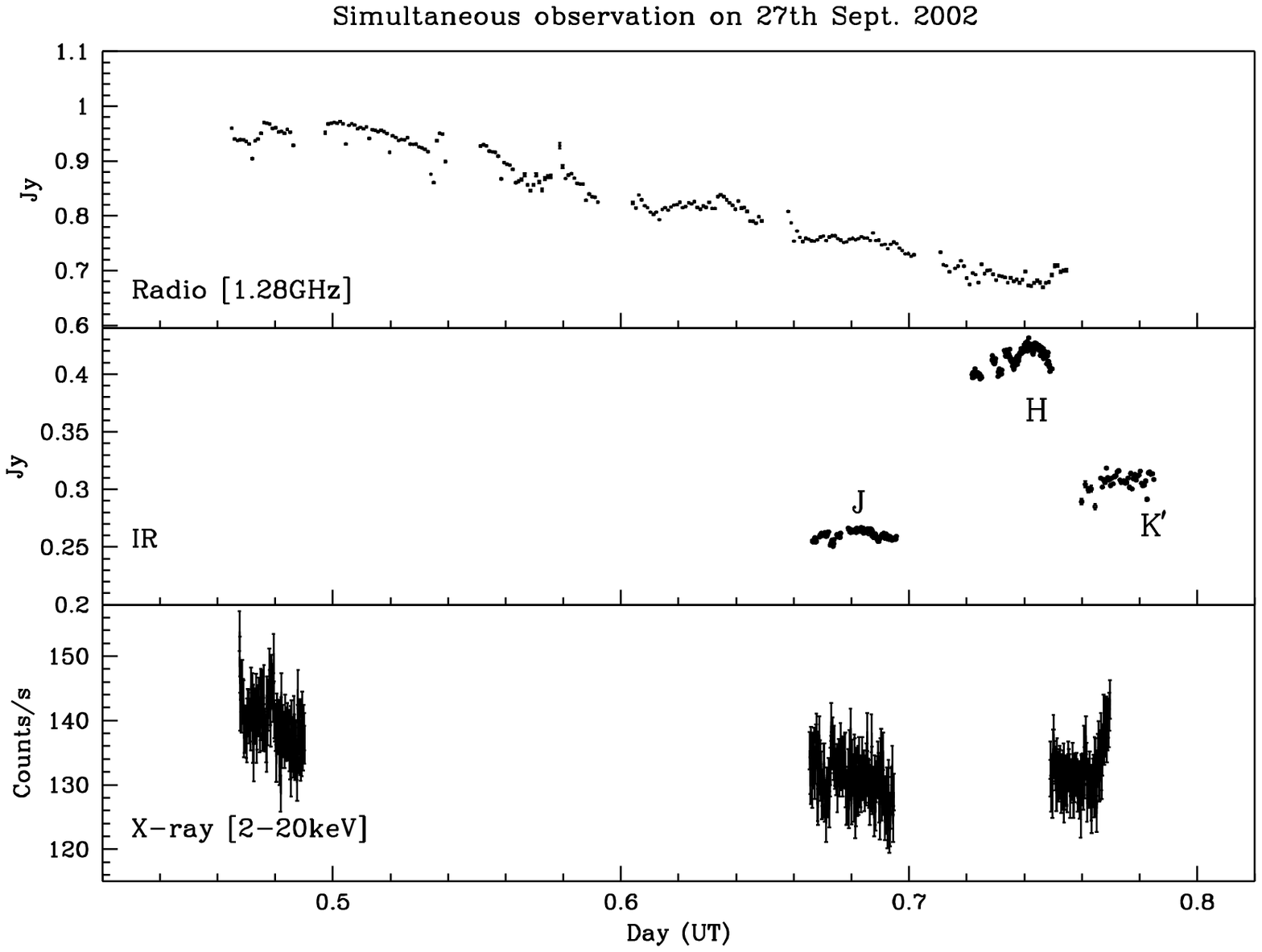}}   
\vskip -3.0cm
{\small{\bf Fig. 1:} Multi wavelength observation of short-time variability
in SS433 by Radio (upper panel), Infra-red (middle panel) nd X-ray (lower panel)
on 27th of September, 2002.}
\end{figure}

The observational result of September 27th, 2002 is shown in Fig. 1
with  UT (Day) along the X-axis. The upper and middle panels show the radio
and IR fluxes (uncorrected for reddening) in Jansky and the lower panel 
shows X-ray counts per second. These observations correspond to an average flux of
$10^{-14}$ergs/cm$^2$/s, $5\times 10^{-10}$ergs/cm$^2$/s and
$ 10^{-10}$ergs/cm$^2$/s respectively. Observations in radio and IR were 
carried out during 25-30th  September, 2002 and no signature of any 
persistent `flare' was observed. The radio data clearly showed a tendency to
go down from $1.0$Jy to $0.7$Jy reaching at about $0.3$Jy on 28th/29th,
while the X-ray data showed a tendency to rise towards the end of
the observation of the 27th. The IR data in each band remained virtually 
constant. The H-band result was found to be higher compared to the J and
K$^\prime$ band results during 25th-29th September, 2002.

Light curves around `local mean' (Fig. 2 of Chakrabarti et al. 2003) clearly 
indicate significant variability in the time-scale of $T_{var} \sim 2-8$ minutes. 
In Fig. 2 the differential flux density variation of IR observations in the J 
and H bands during 27 September 2002 using differential photometry is shown. 
Here we compare the differential flux variation between SS433 and two brightest 
standard stars (std1 and std2) for the whole light curve.  This clearly 
indicates that the variation in the IR light curves of SS433 is intrinsic. 
The analysis shows above 2$\sigma$ level variability in both the bands.

\begin{figure}[t] 
\vskip -3.5 cm
\centerline{\epsfxsize=4.7in\epsfbox{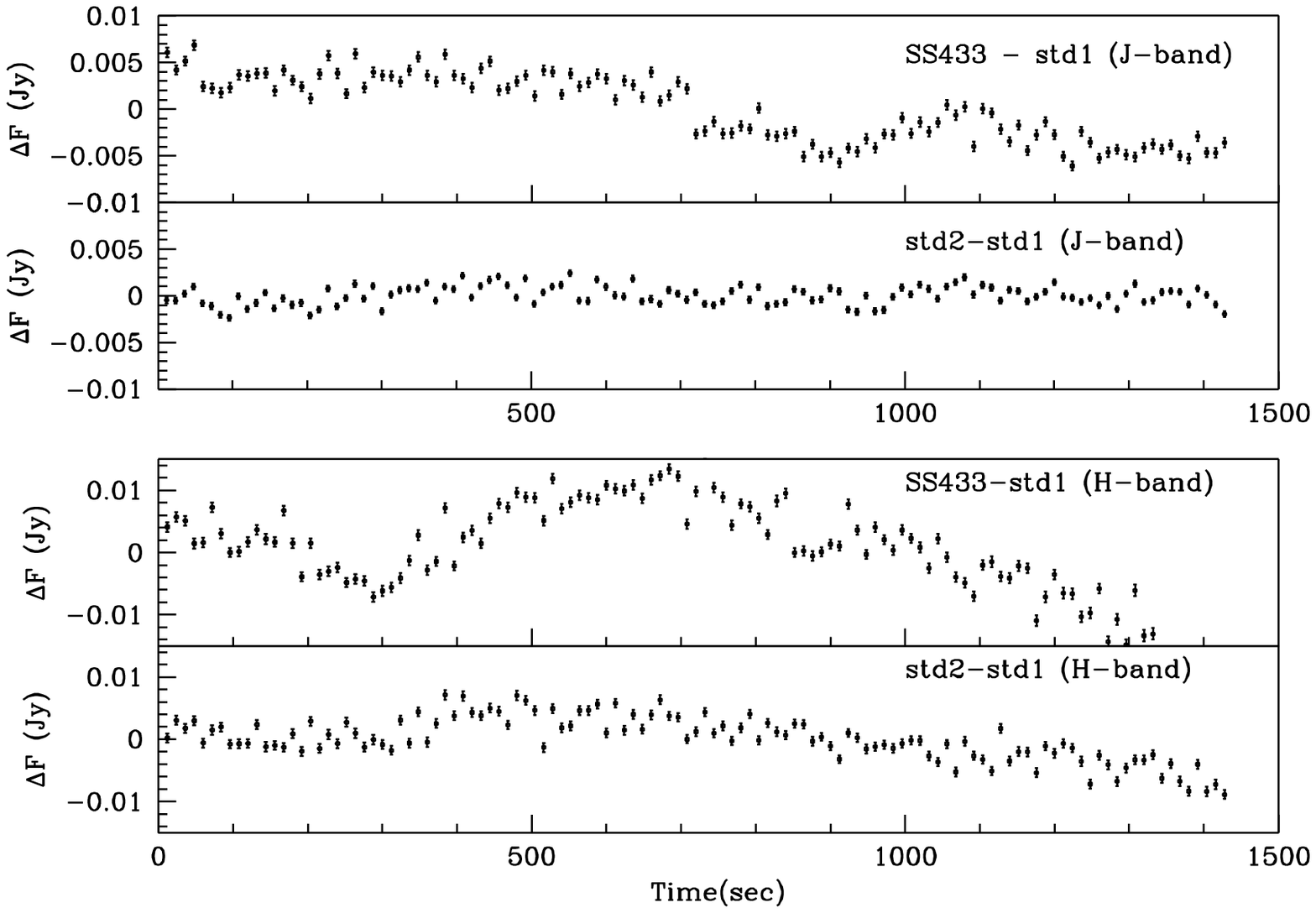}}   
\vskip -0.5cm
{\small{\bf Fig. 2:} 
Differential photometry of SS433 with respect to two brightest standard
stars (std1 and std2) in  the same frame of the object are plotted.
X-axis of the graph is the relative time of measurements in seconds. 
}
\end{figure}

Could these variations be due to individual bullets? In order to be specific,
we present in Fig. 3a, one `micro-flare'-like event in radio from the data
on 29th of Sept., 2002, when radio intensity was further down $\sim 0.3$Jy so
that the micro-flares could be prominently seen. We observe brightening of
the source from $0.35$Jy to $0.8$Jy in $\sim 75$s. The source faded away in 
another $\sim 75$s. Similarly in Fig. 3b, where we presented a `micro-flare'
from the 2nd (central) `spell' of X-ray data of 27th Sept. 2002 (Fig. 1),
we also observe significant brightening  and fading in $\sim 100$s.
The count rate went up more than $15\%$ or so in about a minute.
We calculated the energy contained in the individual radio and X-ray 
micro-flare are $1.1 \times 10^{33}$ergs and $2.7 \times 10^{35}$ergs.
Since the radio luminosity is very small, even when integrated over $0.1$GHz
to $10$GHz radio band (with a spectral index of $\sim -0.5$)
(Vermeulen et al, 1993) we find that almost all the injected energy
at X-ray band is lost on the way during its passage of $\sim 1-2$d.
We also analysed the X-ray spectrum and found that the thermal bremsstrahlung
model with two Fe line features are best fitted. We didn't find any signature
of Keplerian disk in the system.

\begin{figure}[hp] 
\vskip -4.75cm
\centerline{{\epsfxsize=4.7in\epsfbox{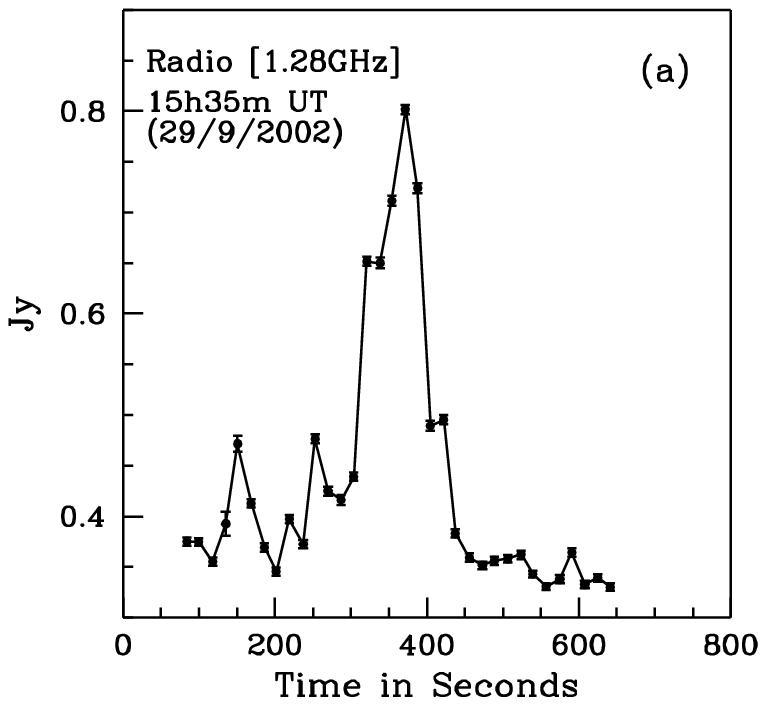}}\hskip -6.0cm {\epsfxsize=4.7in\epsfbox{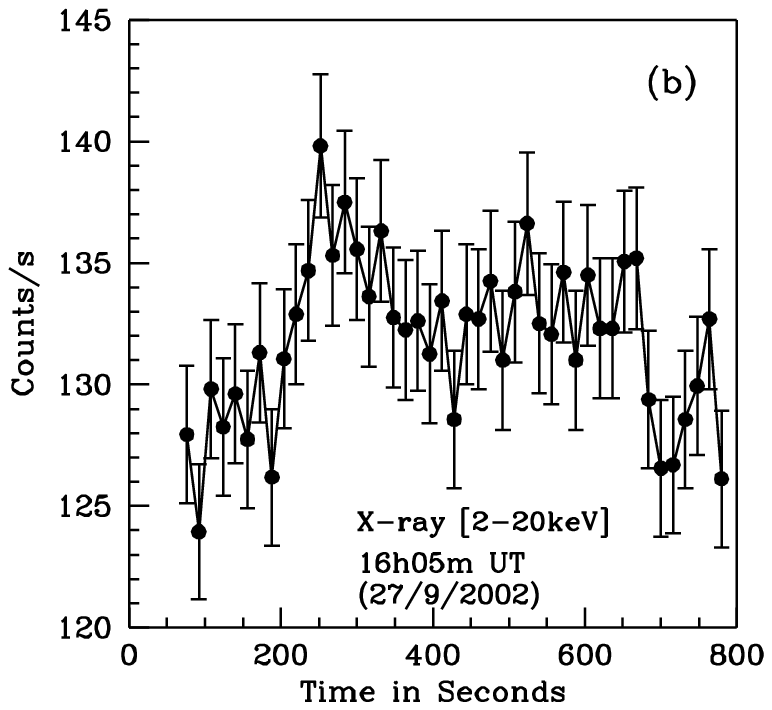}}}
\vskip -2.5cm
{\small{\bf Fig. 3:}
Individual flares in very short time scales are caught. (a) A radio flare 
lasting $2.5$ minutes (observed on 29th Sept. 2002) and (b) an X-ray flare
(observed on 27th of Sept. 2002) lasting for about $3.5$ minutes. Each
bin-size is $16$s}
\end{figure} 

\section{Discussion and conclusions}

Our multi-wavelength observations of X-ray, IR and Radio suggest that
we may be observing the individual bullet like events in different wave
bands. Analysis of the subsequent X-ray observations during inferior and superior conjunctions
on Oct. 2nd, 2003 and March 13th, 2004 respectively confirmed the presence of 
fast variabilities, X-flares, and the fact that the base of the jet is the major source of X-ray
emission (Nandi et al. 2004). We also verified that the Doppler shifts of X-ray lines generally follow
the `kinematic model'.

\section*{Acknowledgments} 
AN acknowledges DST project grant No.  SP/S2/K-15/2001 for funding his research and SP acknowledges
CSIR Fellowship.


\begin{thebibliography}{0}

\bibitem{} Chakrabarti, S.K., Goldoni, P., Wiita, P.J., Nandi, A. \& S. Das, 2002, {\it Astrophys. J.}, {\bf 576}, L45 (2002)

\bibitem{} Chakrabarti, S.K., Pal, S., Nandi, A., Anandarao, B. G. and Mondal, S,
{\it Astrophys. J.}, {\bf 595}, L48 (2003) 

\bibitem{} Margon, B., {\it ARA\&A}, {\bf 22}, 507 (1984)

\bibitem{} Nandi, A., S.K. Chakrabarti, T. Belloni \& P. Goldoni, 2004, MNRAS, submitted

\bibitem{} Vermeulen, R.\ C., et al., {\it Astron. \& Astrophys.}, {\bf 270}, 189 (1993)

\end{thebibliography}
\end{document}